\documentclass[conference]{IEEEtran}
\usepackage[latin9]{inputenc}
\usepackage{multicol,lipsum}
\usepackage{lipsum}
 \usepackage{enumitem}
\usepackage{mathrsfs}
\usepackage{amsmath}
\usepackage{amssymb}
\usepackage{graphicx}
\usepackage{esint}
\usepackage{verbatim}
\usepackage{color}
\usepackage{booktabs}
\usepackage{cite}
\usepackage{amsfonts}
\usepackage{algpseudocode}
\usepackage{algorithm}
\usepackage{epstopdf}

\begin{document}
\title{Decision-Feedback Detection for Bidirectional Molecular Relaying with Direct Links}
\author{
\IEEEauthorblockN{Maryam Khalid and Momin Uppal}
\IEEEauthorblockA{\textit{Department of Electrical Engineering, Lahore University of Management Sciences, Lahore, Pakistan} \\
Emails: mk79@rice.edu,momin.uppal@lums.edu.pk}
}
\maketitle

\begin{abstract} In this paper, we consider bidirectional relaying between two diffusion-based molecular transceivers (bio-nodes). As opposed to existing literature, we incorporate the effect of direct diffusion links between the nodes and leverage it to improve performance. Assuming network coding type operation at the relay, we devise a detection strategy, based on the maximum-likelihood principle, that combines the signal received from the relay and that received from the direct link. At the same time, since a diffusion-based molecular communication channel is characterized by high inter-symbol interference (ISI), we utilize a decision feedback mechanism to mitigate its effect. Simulation results indicate that the proposed setup incorporating the direct link can achieve notable improvement in error performance over conventional detection schemes that do not exploit the direct link and/or do not attempt to mitigate the effect of ISI.
\end{abstract}

\vspace{0.1in}
\begin{IEEEkeywords}
Molecular Communication, Relaying, Maximum Likelihood Detection, Decision feedback equalization
\end{IEEEkeywords}

\IEEEpeerreviewmaketitle

\section{Introduction}
Molecular communication is a naturally occurring phenomenon whereby bio-nodes release chemicals to communicate with receptor nodes in their micro-/nano-scale vicinity. For instance, the human nervous system is composed of nano-networks in which bio-nodes communicate in this fashion \cite{nervous,nano2}. As advancements in nano-technology promise the advent of highly sensitive nano-sized electrical machines \cite{sens1,sens2}, it is envisioned that man-made nano-networks of the future could utilize the bio-inspired molecular communication concept for exchange of information. Thus, unlike conventional wireless or wired communication, these networks may replace electromagnetic signals by molecules as the information bearing entity, with the information encoded in some molecule characteristics. Molecular communication is not only different from conventional communication in terms of its information-transporting nature, but also in terms of its channel characteristics. In particular, the impulse response of a molecular channel, in general, has a long tail that results in high inter-symbol interference (ISI). Moreover, the received noise signal becomes non-stationary and dependent on the transmitted signal. As a result, techniques for encoding, synchronization, estimation, and detection etc. cannot be simply replicated from conventional communication theory. Thus, driven by the envisioned futuristic applications of nano-networks, a fair amount of research into theoretical understanding of molecular communication networks has appeared in recent years.

The performance of molecular communication systems strongly depend on the physical properties of the channel and one such property is the distance between transmitter and receiver. As observed in \cite{att}, a linear increase in distance results in an exponential decrease in concentration of received molecules. Thus, as shown in \cite{JSAC}, the error performance is significantly degraded with an increase in the separation between transmitter-receiver pair. A solution to this problem is deployment of a relay node that helps the transmitter in forwarding its information to the intended receiver; a setup that is referred to as a uni-directional relaying network in conventional communication theory literature. In this paper, we consider the equally important bi-directional relaying network in which two bio-nodes wish to \emph{exchange} their information over a diffusion-based molecular communication channel \cite{bir1,bir2} with the aid of a dedicated relaying node.

\par A fair amount of literature exists on the study of molecular relaying networks; majority of them though have looked at the unidirectional relaying setup. For instance, the design and analysis of relays in calcium signalling channels has been carried out in \cite{ca1,ca2}. Unidirectional relaying for bacteria has been studied in \cite{fekri}, with the communication distance range shown to increase through the use of relays. However, \cite{fekri} does not take ISI into account which is unavoidable in molecular communication channels except at extremely low data-rates. On the other hand, \cite{relay1} incorporates the presence of ISI and derives a mathematical expression for the error probability in a unidirectional molecular decode-forward relaying network. In \cite{AF1}, an expression for the error probability is derived under amplify-and-forward unidirectional molecular relaying. Moreover, the work also attempts to optimize the amplification factor at the relay so as to minimize the error probability. Similarly, the error performance for decode-and-forward molecular unidirectional relaying has been investigated in \cite{DF1}. The work attempts to optimize the decision threshold at the relay so as to minimize the end-to-end error probability. Compared with \cite{AF1}, the work in \cite{DF1} takes drift into account and models the received signal as a Gaussian random variable. As opposed to unidirectional relaying, the only work to have dealt with bidirectional molecular relaying (to the best of the authors' knowledge) is \cite{adnanijaz} in which a fixed-threshold detector is utilized while ignoring the direct links between the transceiver nodes.

As mentioned earlier, this paper considers a bi-directional molecular relaying network. Our work and its contributions over the existing unidirectional relaying works in general, and the bidirectional relaying work \cite{adnanijaz} in particular are summarized in the following.
\begin{itemize}
\item{All existing works cited above ignore the direct paths between the transmitter and the receiver. In many scenarios, as we will show in the subsequent sections, the direct link may be non-negligible. Keeping this in mind, we consider the more general setup in which a given bio-node receives molecules not only from the relay, but also directly from the other transmitting node as well.}
\item{As opposed to \cite{adnanijaz} which utilizes symbol-by-symbol fixed-threshold detection by treating ISI as part of the noise, we utilize a decision-feedback detector at the (non-relaying) nodes; the detector attempts to mitigate the effect of ISI by utilizing decisions of the past.}
\item{We derive an approximate maximum-likelihood (ML) rule for combining the signals from the relay and the direct path, with the proposed rule approximating the transmitted ISI symbols as being equal to the past decisions at the receiver.}
\end{itemize}

\noindent In the presence of a non-negligible direct path between the information bearing nodes, simulation results indicate the significant performance benefit that could be obtained by its exploitation. In addition, the proposed strategy with decision feedback equalization coupled with the ML detection rule is compared with existing fixed-threshold benchmarks, e.g. that of \cite{adnanijaz}, with simulation results indicating clear performance enhancement of the proposed scheme.

The remainder of the paper is organized in the following order. In Section \ref{sec:model}, the bidirectional molecular relaying system model is introduced. This is followed by a description and analysis of the proposed scheme in Section \ref{sec:scheme}. The simulation results are presented in Section \ref{sec:results}, while Section \ref{sec:conclusions} concludes the paper.


\section{System Model} \label{sec:model}
\begin{figure}[h]
\center
\includegraphics[scale=0.52]{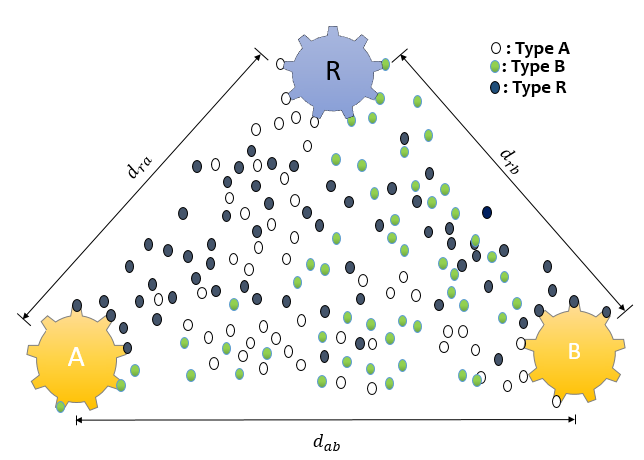}
\caption{Bidirectional molecular relaying setup.} \label{fig:model}
\end{figure}

The bidirectional molecular relaying setup, shown in Fig.~\ref{fig:model}, is composed of three nano-machines, each of which is assumed to be a spherical body of radius-$r$. Two of these machines, referred to as node-$A$ and node-$B$ wish to exchange information while the third machine, referred to as node-$R$, is a dedicated relay that aids this exchange. The distances between $A$ and $B$, $A$ and $R$, and $B$ and $R$ are denoted by $d_{ab}$, $d_{ar}$, and $d_{br}$, respectively, which are assumed to be much greater than the nodes' size. The transmission of information at each node is based on on-off keying, whereby a pulse of $Q$ molecules are released for sending bit 1, while no molecules are released for transmitting the bit 0. We assume that each node is equipped with a distinct molecule type for transmission of information, i.e., $A$, $B$ and $R$ can release distinct Type-$A$, Type-$B$ and \mbox{Type-$R$} molecules, respectively. The molecules from a particular node are assumed to propagate through the medium solely through diffusion without any drift; thus the only mechanism governing the propagation is Brownian motion. Moreover, it is assumed that the molecules do not change their physical properties during propagation and are therefore immutable. Moreover, the nodes are assumed to be perfectly absorbing receivers removing the sensed molecules from the medium. Each node is assumed to be equipped with the receptors for the molecule-types of the other two nodes. Moreover, each node is assumed to operate in the full-duplex mode, i.e., it is assumed to be capable of simultaneously transmitting and receiving \emph{different} molecule types. It is further assumed that all nodes are perfectly synchronized and have complete knowledge of channel behavior, e.g. through the use of training sequences.

\subsubsection{Channel Model}
The  \textit{average} channel impulse response can be derived through the solution of Fick's law that characterizes changes in molecule concentration due to diffusion as a function of time and space. Moreover, the stochastic nature of a molecular channel can be accurately modeled by a Binomial distribution. In particular, if $Q$ molecules are impulsively released at a point-source transmitter, the number of received molecules received at a particular time $t \in [0,T]$ follows a Binomial distribution with a certain success probability $p(t)$. Since, all molecules are independent of each other and their motion is identically distributed, Markovian nature of the system can be exploited to derive the distribution of first hitting time which is then used to compute the success probability $p(t)$. It turns out that the mean of this binomial distribution is equal to the expression derived from Fick's law.

If $Q$ is very large, which it is in practice, a number of approximate models could be utilized. Of the two commonly deployed models, the first is based on approximating the number of molecules hitting the receiver in a particular time interval as a Poisson distribution. The second commonly utilized model, which we employ in this paper, is based on the central limit theorem to approximate the received signal as having a Gaussian distribution. Assuming perfect synchronization between the transmitter and receiver nodes, the discretized signal at the receptors of a given molecule type during symbol time $k$ is given by \cite{JSAC}
\begin{eqnarray}
Y[k] &=& \sum_{j=0}^{L-1}h[j]X[k-j] + Z[k], \nonumber\\
&=& h[0]X[k]  +  \underbrace{\sum_{j=1}^{L-1} h[j] X[k-j]}_{\text{ISI}}+ Z[k], \label{eq:model}
\end{eqnarray}
where $X[k] \in \{0,1\}$ is the transmitted on-off keying symbol sequence of the given molecule type, and $\{h[0], \ldots, h[L-1]\}$ represents the channel impulse response, modeled as an $L$-tap finite impulse response (FIR) filter. The channel coefficients can be well modelled to be obtained from the continuous-time response given by the Fick's law. This response can be modeled by \cite{JSAC,ficklaw},
\begin{equation}\label{eq:Fick}
h(t) = \frac{Q}{\sqrt{(4\pi Dt)^3}}\exp{\left(-\frac{d^2}{4Dt}\right)} u(t)
\end{equation}
where $D$ is the diffusion coefficient, $d$ is the Euclidean distance between the transmitter and the receiver, and $u(t)$ is the unit-step function. It is to be noted that while in principle the channel may have an infinite impulse response, the FIR approximation is justified since the impulse response of a typical molecular channel decays rapidly with time. The term $Z[k]$ is the additive noise which is a manifestation of the molecules' random Brownian motion. As opposed to conventional electromagnetic communication, this noise is signal dependent since the randomness of the received concentration increases proportionally with the number of molecules transmitted. More precisely, if $S[k] = \sum_{j=0}^{L-1} h[j] X[k-j]$ represents the signal part, the noise is modeled as \mbox{$Z[k]\sim \mathcal{N}\left(0,S[k]/\rho\right)$} with $\rho = \frac{4\pi r^3}{3}$ being the receiver volume \cite{JSAC}.





\section{Proposed Detection Scheme for Bidirectional Relaying}\label{sec:scheme}
\begin{figure}[t]
\center
\includegraphics[scale=0.5]{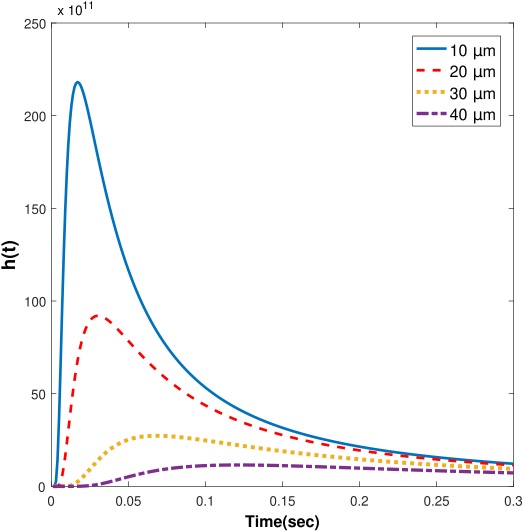}
\caption{Effect of distance on diffusion-based molecular channel response for $D = 2.2 \times 10^{-9}$.} \label{fig:channel}
\end{figure}

In this section, we first describe the bidirectional relaying strategy before discussing the proposed detection scheme. But before that, we take a moment to illustrate the justification behind not neglecting the signal on the direct link. In Fig.~\ref{fig:channel}, we plot the channel impulse response \eqref{eq:Fick} as a function of the distance $d$ between the two transceivers. We observe that although the increase in distance does result in gain attenuation, it does not become completely negligible. For instance, the peak concentration response at a distance of 40 $\mu$m is approximately ten times smaller than that at double the distance of 20 $\mu$m. While this is is indeed small, it cannot be completely ignored. This serves as the primary motivation for us to consider the general case that incorporates the non-negligible albeit weak direct links. As will be illustrated in the next section, incorporating the effect of the direct link provides considerable performance gains.

\subsection{Bidirectional Relaying Strategy}
At any given symbol duration, both nodes $A$ and $B$ release their respective molecule types using the on-off transmission scheme mentioned in Section \ref{sec:model}. Since the relay is equipped with receptors of both molecule types, it has access to two distinct non-interfering received signals $Y_A[k]$ and $Y_B[k]$, which are noisy recordings of the respective transmitted signals and are both of the form given in \eqref{eq:model}. At the relay, we utilize the same methodology as that in \cite{adnanijaz} whereby symbol-by-symbol detection for both users is carried out by treating ISI as part of the noise. In particular, the relay makes a decision $\hat{X}_A[k] = 1$ if $Y_A[k] > \gamma_A$ and $\hat{X}_A[k] = 0$ otherwise, where $\gamma_A$ is a pre-determined threshold. A similar decision $\hat{X}_B[k]$ is made by comparing $Y_B[k]$ with a threshold $\gamma_B$. The message forwarding takes place using a network coding type of an operation. More specifically, the relay transmits the symbol $X_r[k] = \hat{X}_A[k] \oplus \hat{X}_B[k]$ using on-off modulation with \mbox{Type-$R$} molecules, and where $\oplus$ is mod-2 addition. Since the nodes are assumed to operate in the full-duplex mode, the system in the steady state completes the transmission of one symbol per time-slot.

\subsection{Detection Strategy}
Since each node is equipped with two types of receptors, a given node receives two non-interfering signals: one from the relay, and the other directly from the other node. Since the operations at both node-$A$ and node-$B$ are identical, we describe them one of the nodes only. At any of these two nodes, let $Y_d[k]$ represent the signal received directly from the other node. This signal is given as
\vspace{-0.075in}
\begin{equation}\label{eq:Yd}
Y_d[k] = h_d[0]X_d[k] + \sum_{j=1}^{L-1} h_d[j] X_d[k-j] + Z_d[k],
\vspace{-0.05in}
\end{equation}
where $h_d[k]$ is the channel response on the direct link, $Z_d[k]\sim\mathcal{N}\left(0,S_d[k]/\rho\right)$ is the corresponding noise with $S_d[k] = \sum_{j=1}^L h_d[j] X_d[k-j]$, and $X_d[k]$ are the symbols transmitted from the other node (and which need to be recovered). The signal received from the relay is given by
\vspace{-0.075in}
\begin{equation}
Y_r[k] = h_r[0]X_r[k] + \sum_{j=1}^{L-1} h_r[j] X_r[k-j] + Z_r[k],
\vspace{-0.05in}
\end{equation}
where $h_r[k]$ is the channel response on the relay link, while $Z_r[k]$ is the corresponding zero-mean noise with variance proportional to the signal term, as described above. The relay transmitted symbol is obtained as $X_r[k] = \hat{X}_{d,r}[k] \oplus \hat{X}_{0,r}[k]$, where $\hat{X}_{0,r}[k]$ and $\hat{X}_{d,r}[k]$ are the relay's estimates of the receiving node's and the other node's information symbol, respectively. We also note here that in practice, the relay symbol $X_r[k]$ will be a function of $X_d[k-1]$, we assume without loss in generality, that the destination has synchronized itself appropriately to properly align all symbols.

Since the received signal suffers from ISI, its optimum detection requires ML sequence estimation, which is known to  require significant computational complexity. In order to avoid that cost, but still be able to mitigate the effect of ISI, we utilize decision-feedback equalization. In particular, if $\hat{X}_d[k-1], \ldots, \hat{X}_d[k-L+1]$ are the estimates of the previous $L-1$ symbols at the receiving node, we form
\vspace{-0.075in}
\begin{equation}\label{eq:Yd-eq}
\tilde{Y}_d[k] = Y_d[k] - \sum_{j=1}^{L-1}h_d[j] \hat{X}_d[k-j].
\end{equation}
Similarly, in order to mitigate the effect of interference on the relay link, the receiving node forms $\hat{X}_r[i] = \hat{X}_d[i] \oplus X_0[i]$, $i = k-1, \ldots, k-L+1$ (note that the symbol sequence $X_0[k]$ is known perfectly at the receiving node since this is its own data). Using these estimates, it then forms
\vspace{-0.075in}
\begin{equation}\label{eq:Yr-eq}
\tilde{Y}_r[k] = Y_r[k] - \sum_{j=1}^{L-1}h_r[j] \hat{X}_r[k-j].
\end{equation}
Given the equalized outputs \eqref{eq:Yd-eq} and \eqref{eq:Yr-eq}, we attempt to combine them to obtain the decision $\hat{X}_d[k]$. For the purpose, the optimum ML symbol detection rule is obtained by maximizing the joint conditional probability density function (pdf), i.e.
\begin{equation}\label{eq:ML}
\hat{X}_d[k]\! = \! \arg \!  \max_{x_d = 0,1}\! f\!\left(\tilde{Y}_d[k], \tilde{Y}_r[k] \Big{|} X_{d}[k] \!\!=\! x_d, X_0[k] \!\!=\!\! x_0 \right),
\end{equation}
where $x_0$ is the receiver's own data symbol. In order to simplify the computations associated with evaluating the joint pdf in \eqref{eq:ML}, we sub-optimally assume that are no errors in the past decisions $\hat{X}_d[k-1], \ldots, \hat{X}_d[k-L+1]$ as well as $\hat{X}_r[k-1],\ldots,\hat{X}_r[k-L+1]$. As a result, we have
\begin{eqnarray}
\tilde{Y}_d[k] &\approx & h_d[0] X_d[k] + Z_d[k], \text{ and} \label{eq:Yd-approx} \\
\tilde{Y}_r[k] &\approx & h_r[0] X_r[k] + Z_r[k]. \label{eq:Yr-approx}
\end{eqnarray}
Under this assumption, the joint conditional pdf, using the total law of probability can be evaluated as
\begin{equation} \nonumber
f\left(\tilde{Y}_d, \tilde{Y}_r \Big{|} x_d, x_o\right) \!\!=\!\! \sum_{x_r = 0,1} P\left( x_r \Big{|} x_d, x_0\right)
f\left(\tilde{Y}_d, \tilde{Y}_r \Big{|}  x_d, x_r\right),
\end{equation}
where we have omitted the time index for notational convenience. Next, we observe that conditioned on the symbols being transmitted from the relay and the other node, variables $\tilde{Y}_r$ and $\tilde{Y}_d$, based on the approximation in \eqref{eq:Yd-approx} and \eqref{eq:Yr-approx} are independent. Moreover, based on the molecular channel model, they are Gaussian distributed. More precisely, we have
\begin{equation}\label{eq:f}
f\left(\tilde{Y}_d, \tilde{Y}_r \Big{|} x_d, x_r\right) = \frac{e^{- \frac{(\tilde{Y}_d - h_d[0]x_d)^2}{2\sigma_d^2(x_d)} -\frac{(\tilde{Y}_r - h_r[0]x_r)^2}{2\sigma_r^2(x_r)} }}{2\pi \sigma_d(x_d) \sigma_r(x_r)},
\end{equation}
where \vspace{-0.2in}
\[ \sigma_d^2(x_d) = \frac{1}{\rho} [ h_d[0] x_d + \sum_{j=2}^{L-1}h_d[j]\hat{X}_d[k-j]] \text{ and} \]
\[\sigma_r^2(x_r) = \frac{1}{\rho}[h_r[0] x_r + \sum_{j=2}^{L-1}h_r[j]\hat{X}_r[k-j]] \]
are the signal dependent noise variances. We note that the signal variances are not only a function of the current symbol, but also of all the past $L-1$ decisions. However, the notation illustrates explicit dependency only on the current test symbol for the sake of notational ease.

\begin{figure}[t]
\center
\includegraphics[scale=0.35]{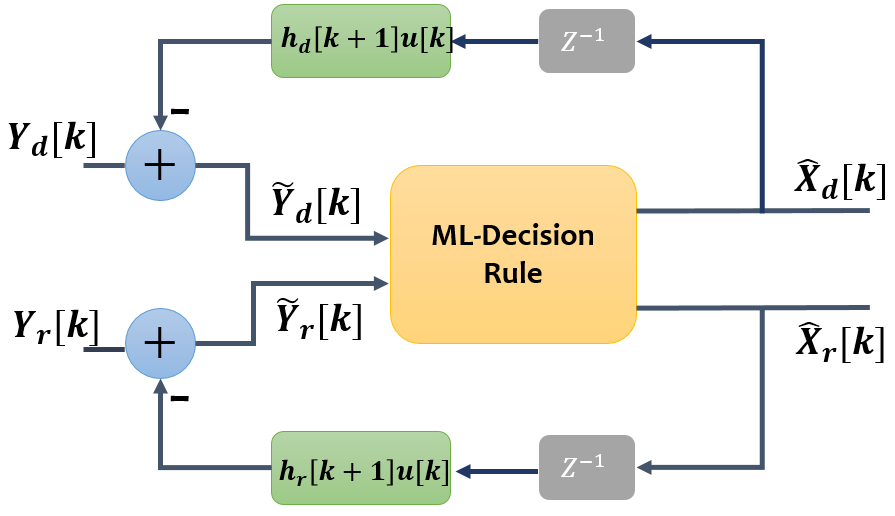}
\caption{Block diagram of the proposed detection strategy.}
\end{figure}


Next, we take a look at evaluating the conditional probability of $X_r$ given $X_d$ and $X_0$. Since $X_r$ is evaluated at the relay as the mod-2 sum of the symbol \emph{decisions}, we need to evaluate the relay's error probability in order to determine the required conditional probability. Recall that relay utilizes symbol-by-symbol detection by comparison of the received signal with fixed thresholds $\gamma_d$ and $\gamma_0$ for the two users; $\gamma_d$ for the user that is to be decoded at the receiving node, and $\gamma_0$ for the receiving node's own symbols. Given the node transmits $X_d[k] = x_d$, the relay received signal is Gaussian with mean and variance given by
\begin{eqnarray}
\mu_{rd}(x_d) &=& h_{rd}[0]x_d + \sum_{j=1}^{L-1} h_{rd}[j] \hat{X}_d[k-j],\\
\sigma_{rd}^2(x_d) &=& \frac{\mu_{rd}(x_d)}{\rho},
\end{eqnarray}
respectively, where the sequence $h_{rd}[0], \ldots, h_{rd}[L-1]$ corresponds to the channel from the transmitting node to the relay. We note that the above is only an approximation, where as before, we have sub-optimally assumed for computational ease that the estimates at the receiver are not erroneous. In addition, we once again note that even though these parameters are a function of $k$ and the estimates of the past, we do not explicitly illustrate this dependence for notational ease. Now, for $x_d = 0,1$, let
\begin{equation}
g_{d}\left(x_d\right) \triangleq Q\left( \frac{\gamma_d - \mu_{rd}(x_d)}{\sigma_{rd}(x_d)}\right),
\end{equation}
where $Q(x) = \frac{1}{\sqrt{2\pi}} \int_{x}^\infty e^{-\frac{x^2}{2}} dx $ is the tail probability of a standard normal random variable. Then it is easy to see that the (approximate) probability that the relay makes an erroneous decision when $X_d = x_d$ was transmitted is given by
\begin{equation}
P_{e,d}\left(x_d\right) \triangleq \left\{ \begin{array}{ll}g_d\left(x_d\right)&x_d = 0 \\1 - g_d\left(x_d\right)&x_d = 1  \end{array} \right.
\end{equation}
Similarly, we define \vspace{-0.1in}
\begin{eqnarray}
\mu_{r0}(x_0) &=& h_{r0}[0]x_0 + \sum_{j=1}^{L-1} h_{r0}[j] {X}_0[k-j],\\
\sigma_{r0}^2(x_0) &=& \frac{\mu_{r0}(x_0)}{\rho},
\end{eqnarray}
as the (exact) mean and variance of the relay signal received from the decoding node's own transmissions. Similarly, for $x_0 = 0,1$, let
\begin{equation}
g_{0}\left(x_0\right) \triangleq Q\left( \frac{\gamma_0 - \mu_{r0}(x_0)}{\sigma_{r0}(x_0)}\right).
\end{equation}
Then the probability that the relay makes an error in its decision of $X_0$ when $X_0 = x_0$ was transmitted is given by
\begin{equation}
P_{e,0}\left(x_0\right) \triangleq \left\{ \begin{array}{ll}g_0\left(x_0\right)&x_0 = 0 \\1 - g_0\left(x_0\right)&x_0 = 1  \end{array} \right.
\end{equation}

Given the definitions above, we are now ready to determine the conditional probability $P\left(x_r | x_d,x_0\right)$ required for the ML decision in \eqref{eq:ML}. Recall that the relay utilizes network coding operation with $X_r = \hat{X}_d \oplus \hat{X}_0$. Thus if $x_r = x_d \oplus x_0$, it implies the event that either the relay estimated both symbols correctly or both were estimated in error. On the other hand $x_r \neq x_d \oplus x_0$ implies that exactly one of the two symbols was detected in error at the relay. Thus

\noindent $P\left(x_r \Big{|} x_d,x_0\right)=$
\begin{equation}
\!\left\{ \!\!\!\begin{array}{ll} P_{e,d}\left(\!x_d\!\right)\!P_{e,0}\left(\!x_0\!\right) \!+\! \overline{P_{e,d}}\left(\!x_d\!\right)\!\overline{P_{e,0}}\left(\!x_0\!\right) & x_r = x_d \oplus x_0 \\
P_{e,d}\left(\!x_d\!\right)\!\overline{P_{e,0}}\left(\!x_0\!\right) \!+\! \overline{P_{e,d}}\left(\!x_d\!\right)\!P_{e,0}\left(\!x_0\!\right) & x_r \neq x_d \oplus x_0, \end{array} \right. \label{eq:Pxr}
\end{equation}
with $\overline{P_{e,d}}\left(x_d\right) \triangleq 1 - P_{e,d}\left(x_d\right)$ and $\overline{P_{e,0}}\left(x_0\right) \triangleq 1 - P_{e,0}\left(x_0\right)$. Substituting \eqref{eq:f} and \eqref{eq:Pxr} in \eqref{eq:ML} yields the approximate decision rule, which for $x_0 = 0$ can be rearranged to obtain the following compact form (a similar rule can be written for $x_0 = 1$, but is not presented here due to space limitations)
\begin{equation}
\label{eq:avg}
w_1 \Phi(\tilde{Y}_d) + w_2 \Phi(\tilde{Y}_d)\Psi( \tilde{Y}_r) + w_3 \Psi(\tilde{Y}_r) \lessgtr^1_0 \beta
\end{equation}
with
\begin{eqnarray}
\Phi (\tilde{Y}_d) &=&  e^{- \frac{(\tilde{Y}_d-h_d[0])^2}{2\sigma_d^2(1)} + \frac{\tilde{Y}_d^2}{2\sigma_d^2(0)}}\\
\Psi (\tilde{Y}_r) = e^{- \frac{(\tilde{Y}_r-h_r[0])^2}{2\sigma_r^2(1)} + \frac{\tilde{Y}_r^2}{2\sigma_r^2(0)}}
\end{eqnarray} 
being non-linear exponential-type functions of the (equivalent) received signals $\tilde{Y}_d$ and $\tilde{Y}_r$. Also, $w_1$, $w_2$, and $w_3$ are the weights that are a function of the error probabilities $P_{e,d}(\cdot)$ and $P_{e,0}(\cdot)$ and are given by
\begin{eqnarray}
w_1 &=& \frac{\sigma_d(0)}{\sigma_d(1)}\left[P_{e,d}\left(1\right)\overline{P_{e,0}}\left(0\right) + \overline{P_{e,d}}\left(1\right)P_{e,0}\left(0\right) \right], \nonumber \\
w_2 &=& \frac{\sigma_d(0)\sigma_r(0)}{\sigma_d(1)\sigma_r(1)}\left[ P_{e,d}\left(1\right)P_{e,0}\left(0\right) + \overline{P_{e,d}}\left(1\right)\overline{P_{e,0}}\left(0\right)\right], \nonumber \\
w_3 &=& \frac{\sigma_r(0)}{\sigma_r(1)} \left[P_{e,d}\left(0\right)\overline{P_{e,0}}\left(0\right) + \overline{P_{e,d}}\left(0\right)P_{e,0}\left(0\right) \right]. \nonumber \\
\end{eqnarray}
Finally, $\beta$ is the threshold given by
\begin{equation}
\beta = \left[ P_{e,d}\left(0\right)P_{e,0}\left(0\right) + \overline{P_{e,d}}\left(0\right)\overline{P_{e,0}}\left(0\right)\right].
\end{equation}
In short, the decision metric is a non-linear weighted average of the received signals, which is then compared with a threshold to decide whether $x_d = 0$ or $x_d = 1$ was transmitted. As one would expect, it can be verified that if the likelihood of relay making an error is high, the signal from direct path is given a higher weight. On the other hand, when the error probability at relay is low, more significance is given to relay link in making the decision.


\section{Simulation Results}\label{sec:results}
In order to compare the performance of our detector with other benchmarks, we use the end-to-end bit-error rate (BER) as a performance measure. Using the same definition as in \cite{JSAC}, we consider the signal-to-noise ratio on a link to be given by
\begin{equation}
 \label{eq:SNR}
\text{SNR} = {\frac{1}{L}\sum_{k=0}^{L-1} |h[k]|^2}\left({\frac{0.5}{\rho}\sum_{k=0}^{L-1} h[k]}\right)^{-1}
\end{equation}
Since the SNR is a function of the channel coefficients only, it is increased on a given link by increasing the number of molecules transmitted.
For the purpose of simulations, we fix the diffusion coefficients for Type-$A$, Type-$B$ and Type-$R$ molecules to be $6 \times 10^{-9}$, $5 \times 10^{-9}$ and $4.3 \times 10^{-9}$ $m^2/s$, respectively. The respective distances are set to $d_{ab} = 2$ $\mu m$, $d_{ar} = 1$ $\mu m$, $d_{br} = 1$ $\mu m$. Using these parameters, the channel impulse response on each link was generated using Fick's law \eqref{eq:Fick}, and the received signal model of \eqref{eq:model} was simulated.

For performance comparison, we use two benchmarks. The first benchmark is a fixed-threshold detector, similar to that of \cite{adnanijaz}, which neither incorporates the effect of the direct path, nor does it try to mitigate the effect of ISI. For our simulations, we fix this threshold to be equal to $h_r[0]/2$ (half of the first channel tap on the relay link). The second benchmark utilizes a decision-feedback equalizer, similar to that in \cite{JSAC}, once again \emph{without} exploiting the direct link. For this benchmark too, after cancelling the effect of past estimated symbols, the threshold used was equal to $h_r[0]/2$. For both the benchmarks, as well as the proposed scheme, the detection threshold at the relay was chosen once again to be half of the first channel tap. In Fig.(4), we plot the BER of the proposed scheme (and the benchmarks) as a function of the SNR.

\begin{figure}
\includegraphics[scale=0.5]{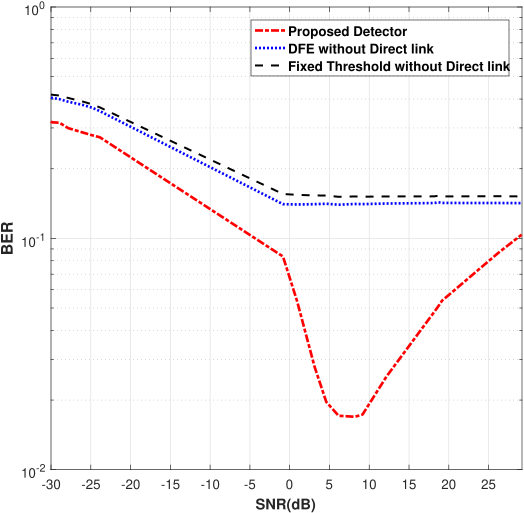}
\caption{Comparison between proposed ML-based detector and threshold detectors for $L_h=10$ and Node radii $r = 80nm$}
\end{figure}

\begin{figure}

\includegraphics[scale=0.55]{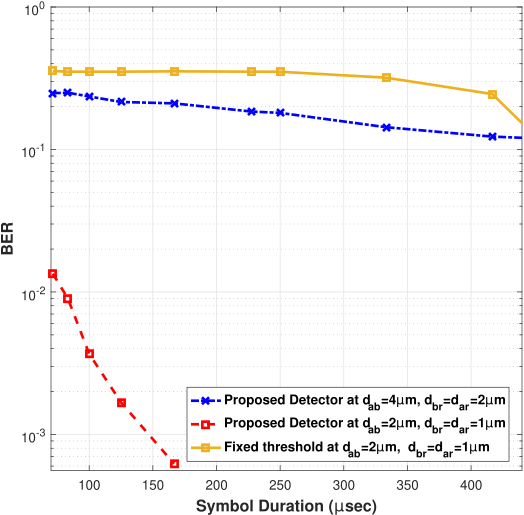}
\caption{Impact of ISI on error Performance}
\vspace{-0.5cm}
\end{figure}

We see that under the system parameters described above, the performance of the fixed-threshold detector does not improve much with increasing SNR due to the effect of significant amount of ISI. Similarly, we see that while the second benchmark is better than the fixed-threshold detector, the proposed scheme clearly has a significant performance benefit over the two benchmarks.  The main reasons for this are (a) decision feedback equalization to mitigate the effect of ISI, (b) incorporating the direct path in the decision process, and (c) the (approximate) ML detection strategy. It is important to point out that the BER performance, as can be seen in the figure, is not monotonically decreasing with SNR. This can be explained by the fact that an at very  high values of SNR and essentially at higher Q, the performance suffers from higher error propagation through the decision-feedback mechanism. In Fig(5), the impact in symbol duration on the performance of proposed strategy is observed. It can be seen that as symbol duration increases and therefore ISI decreases, the error probability decreases. Keeping the fact in mind that error performance of bidirectional setup is poor compared to unidirectional setup, the results presented above are consistent with those reported in \cite{adnanijaz}\cite{unconv} and the error probability around $10^{-2}$ is acceptable in relay-aided long range Molecular communication systems\cite{AF1}\cite{fekri}. Hence, the proposed detector is robust to not only brownian noise but also to ISI, both of which are inherent features of molecular channel.


\section{Conclusions}\label{sec:conclusions}
We have proposed a detection strategy for bidirectional molecular relaying while incorporating the effect of the direct links. The performance of the proposed scheme is found to have considerable performance benefits over existing benchmarks. Future extensions include further improvements in detector design that incorporates soft-ISI removal instead of the hard cancellation currently employed. Further work is also required to find analytical expressions for the probability of error in bidirectional molecular relaying with direct links.

\bibliographystyle{IEEEtran}  

\bibliography{main}  

\begin{thebibliography}{10}
\providecommand{\url}[1]{#1}
\csname url@samestyle\endcsname
\providecommand{\newblock}{\relax}
\providecommand{\bibinfo}[2]{#2}
\providecommand{\BIBentrySTDinterwordspacing}{\spaceskip=0pt\relax}
\providecommand{\BIBentryALTinterwordstretchfactor}{4}
\providecommand{\BIBentryALTinterwordspacing}{\spaceskip=\fontdimen2\font plus
\BIBentryALTinterwordstretchfactor\fontdimen3\font minus
  \fontdimen4\font\relax}
\providecommand{\BIBforeignlanguage}[2]{{%
\expandafter\ifx\csname l@#1\endcsname\relax
\typeout{** WARNING: IEEEtran.bst: No hyphenation pattern has been}%
\typeout{** loaded for the language `#1'. Using the pattern for}%
\typeout{** the default language instead.}%
\else
\language=\csname l@#1\endcsname
\fi
#2}}
\providecommand{\BIBdecl}{\relax}
\BIBdecl

\bibitem{nervous}
D.~Malak and O.~Akan, ``Communication theoretical understanding of intra-body
  nervous nanonetworks,'' \emph{IEEE Communications Magazine}, vol.~52, no.~4,
  pp. 129--135, 2014.

\bibitem{nano2}
D.~Malak and O.~B. Akan, ``Molecular communication nanonetworks inside human
  body,'' \emph{Nano Communication Networks}, vol.~3, no.~1, pp. 19--35, 2012.

\bibitem{sens1}
M.~Curreli, R.~Zhang, F.~N. Ishikawa, H.-K. Chang, R.~J. Cote, C.~Zhou, and
  M.~E. Thompson, ``Real-time, label-free detection of biological entities
  using nanowire-based fets,'' \emph{IEEE Transactions on Nanotechnology},
  vol.~7, no.~6, pp. 651--667, 2008.

\bibitem{sens2}
Z.~Li, Y.~Chen, X.~Li, T.~Kamins, K.~Nauka, and R.~S. Williams,
  ``Sequence-specific label-free dna sensors based on silicon nanowires,''
  \emph{Nano Letters}, vol.~4, no.~2, pp. 245--247, 2004.

\bibitem{att}
I.~Llatser, A.~Cabellos-Aparicio, M.~Pierobon, and E.~Alarc{\'o}n, ``Detection
  techniques for diffusion-based molecular communication,'' \emph{IEEE Journal
  on Selected Areas in Communications}, vol.~31, no.~12, pp. 726--734, 2013.

\bibitem{JSAC}
D.~Kilinc and O.~B. Akan, ``Receiver design for molecular communication,''
  \emph{IEEE Journal on Selected Areas in Communications}, vol.~31, no.~12, pp.
  705--714, December 2013.

\bibitem{bir1}
P.~Larsson, N.~Johansson, and K.~E. Sunell, ``Coded bi-directional relaying,''
  in \emph{2006 IEEE 63rd Vehicular Technology Conference}, vol.~2, May 2006,
  pp. 851--855.

\bibitem{bir2}
S.~Katti, H.~Rahul, W.~Hu, D.~Katabi, M.~Medard, and J.~Crowcroft, ``Xors in
  the air: Practical wireless network coding,'' \emph{IEEE/ACM Transactions on
  Networking}, vol.~16, no.~3, pp. 497--510, June 2008.

\bibitem{ca1}
T.~Nakano and J.-Q. Liu, ``Design and analysis of molecular relay channels: An
  information theoretic approach,'' \emph{IEEE Transactions on NanoBioscience},
  vol.~9, no.~3, pp. 213--221, 2010.

\bibitem{ca2}
T.~Nakano and J.~Shuai, ``Repeater design and modeling for molecular
  communication networks,'' in \emph{2011 IEEE Conference on Computer
  Communications Workshops (INFOCOM WKSHPS)}, April 2011, pp. 501--506.

\bibitem{fekri}
A.~Einolghozati, M.~Sardari, and F.~Fekri, ``Decode and forward relaying in
  diffusion-based molecular communication between two populations of biological
  agents,'' in \emph{Communications (ICC), 2014 IEEE International Conference
  on}.\hskip 1em plus 0.5em minus 0.4em\relax IEEE, 2014, pp. 3975--3980.

\bibitem{relay1}
X.~Wang, M.~D. Higgins, and M.~S. Leeson, ``Relay analysis in molecular
  communications with time-dependent concentration,'' \emph{IEEE Communications
  Letters}, vol.~19, no.~11, pp. 1977--1980, 2015.

\bibitem{AF1}
A.~Ahmadzadeh, A.~Noel, A.~Burkovski, and R.~Schober, ``Amplify-and-forward
  relaying in two-hop diffusion-based molecular communication networks,'' in
  \emph{Global Communications Conference (GLOBECOM), 2015 IEEE}.\hskip 1em plus
  0.5em minus 0.4em\relax IEEE, 2015, pp. 1--7.

\bibitem{DF1}
N.~Tavakkoli, P.~Azmi, and N.~Mokari, ``Performance evaluation and optimal
  detection of relay-assisted diffusion-based molecular communication with
  drift,'' \emph{IEEE transactions on nanobioscience}, vol.~16, no.~1, pp.
  34--42, 2017.

\bibitem{adnanijaz}
A.~Aijaz, A.~H. Aghvami, and M.~R. Nakhai, ``On error performance of network
  coding in diffusion-based molecular nanonetworks,'' \emph{IEEE Transactions
  on Nanotechnology}, vol.~13, no.~5, pp. 871--874, 2014.

\bibitem{ficklaw}
H.~B. Yilmaz, A.~C. Heren, T.~Tugcu, and C.~B. Chae, ``Three-dimensional
  channel characteristics for molecular communications with an absorbing
  receiver,'' \emph{IEEE Communications Letters}, vol.~18, no.~6, pp. 929--932,
  June 2014.

\bibitem{unconv}
A.~Ahmadzadeh, A.~Noel, and R.~Schober, ``Analysis and design of multi-hop
  diffusion-based molecular communication networks,'' \emph{IEEE Transactions
  on Molecular, Biological and Multi-Scale Communications}, vol.~1, no.~2, pp.
  144--157, June 2015.

\end{thebibliography}





\end{document}